\documentclass[graybox]{svmult}

\usepackage{mathptmx}       
\usepackage{helvet}         
\usepackage{courier}        
\usepackage{makeidx}         
\usepackage{graphicx}        
\usepackage{multicol}        

\makeindex             

\usepackage{amsmath}
\usepackage{amssymb}
\usepackage{amsfonts}

\newcommand{\ep}{\varepsilon}

\newcommand{\ffi}{\varphi}

\newcommand{\R}{\mathbb{R}}

\newcommand{\PP}{\mathbb{P}}
\newcommand{\EE}{\mathbb{E}}

\newcommand{\T}{\mathcal{T}}
\newcommand{\LL}{\mathcal{L}}

\newcommand{\mbf}{\mathbf}

\newcommand{\h}{\tilde h_\ep}

\begin{document}

\title*{Diffusive limit for the random Lorentz gas}
\author{Alessia Nota}
\institute{Alessia Nota \at Dipartimento di Matematica, Universit\`a di Roma La Sapienza\\ 
Piazzale Aldo Moro 5, 00185 Roma -- Italy ,
 \email{nota@mat.uniroma1.it}}
%
%
\maketitle

\abstract{
We review some recent results concerning the derivation of the diffusion equation and the validation of Fick's law for the microscopic model given by the random Lorentz Gas. These results are achieved by using a linear kinetic equation 
as an intermediate level of description between our original mechanical system and the diffusion equation. }


\section{Introduction}\label{sec1}
The problem of deriving
macroscopic evolution equations from the microscopic laws of
motion governed by Newton's laws of classical mechanics is one of the most important keystones in mathematical physics. 

Here we consider a simple microscopic model, namely a gas of non-interacting particles in a fixed random configuration of scatterers. This dynamical system is usually referred to as the Lorentz gas, since it was proposed by H. A. Lorentz in 1905, see \cite{L}, to explain the motion of electrons in metals applying the methods of the kinetic theory of gases. Even though this model is quite simple, it is still paradigmatic. Indeed complexities and interesting features come up in the analysis showing new and unexpected macroscopic phenomena.

The Lorentz gas consists of a particle moving through infinitely heavy, randomly distributed scatterers. The interaction between the Lorentz particle and the scatterers is specified by a central potential of finite range.  Hence the motion of the Lorentz particle is defined through the solution of Newton's equation of motion.
Lorentz's idea was to view electrons as a gas of light particles colliding with the metallic atoms; neglecting collisions between electrons, Lorentz described the interaction of electrons with the metallic atoms by a collision integral analogous to Boltzmann's. The original system is Hamiltonian, the only stochasticity being that of 
the positions of the scatterers. This randomness is absolutely necessary to obtain the correct kinetic description. Indeed, for this system, one can prove, under suitable scaling limits, a rigorous validation of linear kinetic equations and, from this, of diffusion equations. 

We can argue in terms of stochastic processes. The motion of the Lorentz particle is a stochastic process which is non Markovian. The scaling limit procedure can be understood as a Markovian approximation which leads to a Markov process whose forward equation is a suitable kinetic equation. More precisely the scaling limits we are considering consist of a kinetic scaling of space and time, namely $t\to \ep t$, $x\to \ep x$ and a suitable rescaling of the density of the obstacles and the intensity of the interaction.
Accordingly to the resulting frequency of collisions, the mean free path of the particle can have or not macroscopic length and different kinetic equations arise. Typical examples are the linear Boltzmann equation and the linear Landau equation.

The first scaling one could consider is the Boltzmann-Grad limit. The first result in this direction was obtained by Gallavotti in 1969, see \cite{G}, who derived the linear Boltzmann equation starting from 
a random distribution of fixed hard scatterers in the Boltzmann-Grad limit (low density), namely when the number of collisions is small, thus the mean free path of 
the particle is macroscopic. This result was improved and extended to more general distribution by Spohn \cite{S}. In \cite{BBS} Boldrighini, Bunimovich and Sinai proved that the limiting Boltzmann equation holds for almost every scatterer configuration drawn from a Poisson distribution. Moreover, for the sake of completeness, we refer to \cite{GS-RT}, \cite{PSS} for a rigorous derivation of the nonlinear Boltzmann equation from a system of hard spheres, or a system of Newtonian particles interacting via a short-range potential, in the low density limit.
As we already pointed out we remind that the randomness of the distribution of the scatterers is essential in the derivation of the linear Boltzmann equation, in fact for a periodic configuration of scatterers the linear Boltzmann equation fails (see \cite{CG1}), and the random flight process that emerges in the Boltzmann-Grad limit is substantially more complicated. The first complete proof of the Boltzmann-Grad limit of the periodic Lorentz gas, valid for all lattices and in all space dimensions, can be found in \cite{MS}. The mathematical properties of the generalized linear Boltzmann equation derived are analyzed in \cite{CG2}.

Another scaling of interest is the weak coupling limit. The idea of the weak coupling limit is that, by some kind of central limit effect, very many but weak collisions should lead to a diffusion type evolution. The correct kinetic equation which is derived in this scaling limit is the Linear Landau equation
\begin{equation}\label{Landau}
(\partial_{t}+v\cdot\nabla_{x})f(x,v,t)=B \Delta_{|v|} f(x,v,t),
\end{equation}
where $\Delta_{|v|}$ is the Laplace-Beltrami operator on the $d$-dimensional sphere 
of radius $|v|$. It is a Fokker-Planck equation for the stochastic process $(V(t),X(t))$, where the velocity process $V$ is a Brownian motion on the (kinetic) energy sphere, and the position $X$ is an additive functional of $V$. The velocity diffusion follows from the facts that 
there are many elastic  collisions. 
The diffusion coefficient $B$ is proportional to the variance of the transferred momentum in a single collision and depends on the shape of the interaction potential. The first  result in this direction was obtained by
Kesten and Papanicolau for a particle in $\R^3$ in a weak mean zero random force field, see \cite{KP}. D\"urr, Goldstein and Lebowitz proved that in $\R^2$ the velocity process converges in distribution to Brownian motion on a surface of constant speed for sufficiently smooth interaction potentials, see \cite{DGL}. 

The linear Landau equation appears also in an intermediate scale between the low density and the weak-coupling regime, namely when the (smooth) interaction potential $\phi$ rescales according to 
$\phi\to \ep^\alpha \phi$, $\alpha\in (0,1/2)$ 
and  the density of the obstacles is of order $\ep^{-2\alpha-(d-1)}$ (\cite{DR}, \cite{K}). 
The limiting cases $\alpha=0$ and $\alpha=1/2$ correspond respectively to the low density limit and the weak-coupling limit.

The rigorous derivation of hydrodynamical equations grounds on the heuristic idea that after a few mean free times the Lorentz gas is already very close to the local equilibrium which subsequently evolves according to the diffusion equation.
Clearly the only hydrodynamic equation for the Lorentz gas is the diffusion equation since the only conserved quantity is the mass. 

The rigorous derivation of the heat equation from the mechanical system given by the Lorentz gas is actually a very difficult and still unsolved problem. In fact we would expect that, under the diffusive scaling, the distribution density of the test particle converges to that of a diffusion process.
Bunimovich and Sinai (see \cite{BS}) showed that such diffusive limit holds when the scatterers are periodically distributed. This is the most important result in the transition from the microscopic to the macroscopic description.  
 
Nonetheless 
one can handle this problem by deriving the diffusion equation from the correct kinetic equation which arises, according to the suitable kinetic scaling performed, from the random Lorentz gas. 
We remark, however, that the hydrodynamics for the Lorentz model is not equivalent to the hydrodynamics for the kinetic equation. 

In this direction, in \cite{BNP}, we provide a rigorous derivation of the heat equation from the particle system (the Lorentz model) using the linear Landau equation as a bridge between our original mechanical system and the diffusion equation. It works once having an explicit control of the error in the kinetic limit (see also \cite{DP}, where the set of bad configurations are explicitly estimated). The diffusive limit can be achieved since the control of memory effects still holds for a longer time scale.

Moreover, since it is well known how important and challenging is the characterization of stationary nonequilibrium states exhibiting transport  phenomena in the rigorous approach to nonequilibrium Statistical Mechanics, we are interested in considering the Lorentz model out of equilibrium. 
Energy or mass transport in non equilibrium macroscopic systems are described phenomenologically by Fourier's  and Fick's law respectively. There are very few rigorous results in this direction in the current literature (see for instance \cite{LS}, \cite{LS1}, \cite{LS2}).
A contribution in this direction, discussed in Section \ref{sec:Persp}, is the validation of the Fick's law for the Lorentz model in a low density situation which has been recently proven in \cite{BNPP}. 

\section{From Microscopic to Macroscopic Description}
We consider a Poisson distribution of fixed scatterers in $\R^2$ and denote by $c_1,\dots, c_N$ their centers. This means that, given $\mu>0$, the probability density of finding $N$ obstacles in a bounded measurable set $A\subset \R^2$ is 
\begin{equation}\label{eq:prob}
\PP(d\textbf{c}_N)=e^{-\mu |A|}\frac{\mu^{N}}{N!}dc_1\dots dc_N,
\end{equation}
where $|A|=\text{meas}(A)$ and  $\textbf{c}_N=(c_1,\dots,c_N).$
The equations of motion for the point particle of unitary mass are
\begin{equation}\label{eqmot}
\left\{\begin{array}{ll}
\dot{x}=v&\\
\dot{v}=-\sum_{i=1}^{N}\nabla\phi(|x-c_i|)&,
\end{array}\right.
\end{equation} 
where $x$ and $v$ denote position and velocity of the test particle, $t$ the time and, as usual, $\dot{A}=\frac{\,dA}{\,dt}$ indicates the time derivative for any time dependent variable $A$. Finally $\phi:\R^{+}\to\R$ is given by
\begin{equation}
\label{pot barr}
\phi( r)=\left\{\begin{array}{ll}
1\quad \text{if}\; r<1&\\
0\quad \text{otherwise}&,
\end{array}\right.
\end{equation}
namely a circular potential barrier. 

This choice for the potential arises from a problem of geometric optics. We are looking at the optical path followed by a light ray traveling in a inhomogeneous medium. More precisely we have a medium, for example water, in which circular drops of a different substance are distributed. These drops are made of a different substance with smaller refractive index, for example air.
The analogy between geometric optics and classical mechanics implies that the trajectory of the light ray is the trajectory of a test particle moving in a random distribution of scatterers where each scatterer generates a circular potential barrier.

To outline a kinetic behavior of the particle, we introduce the scale parameter $\ep>0$, indicating the ratio between the macroscopic and the microscopic variables, and rescale according to 
\begin{equation}\label{scaling}
x\rightarrow\ep x,\; t\rightarrow\ep t,\; \phi\rightarrow\ep^{\alpha}\phi
\end{equation}
with $\alpha\in[0,1/2]$. Then Eq.ns \eqref{eqmot} become
\begin{equation}\label{scaled}
\left\{\begin{array}{ll}
\dot{x}=v&\\
\dot{v}=-\ep^{\alpha-1}\sum_{i}\nabla\phi(\frac{|x-c_i|}{\ep})&.
\end{array}\right.
\end{equation}
 We rescale also the intensity $\mu$ of the scatterers as $\mu_{\varepsilon}=\mu\varepsilon^{-\delta}$, where $\delta=1+2\alpha$. Accordingly we denote by $\PP_{\ep}$ the probability density \eqref{eq:prob} with $\mu$ replaced by $\mu_{\ep}$ and $\EE_{\ep}$ will be the expectation with respect to the measure $\PP_{\ep}.$
 
\begin{figure}[ht]
\centering
\includegraphics[scale= 0.34]{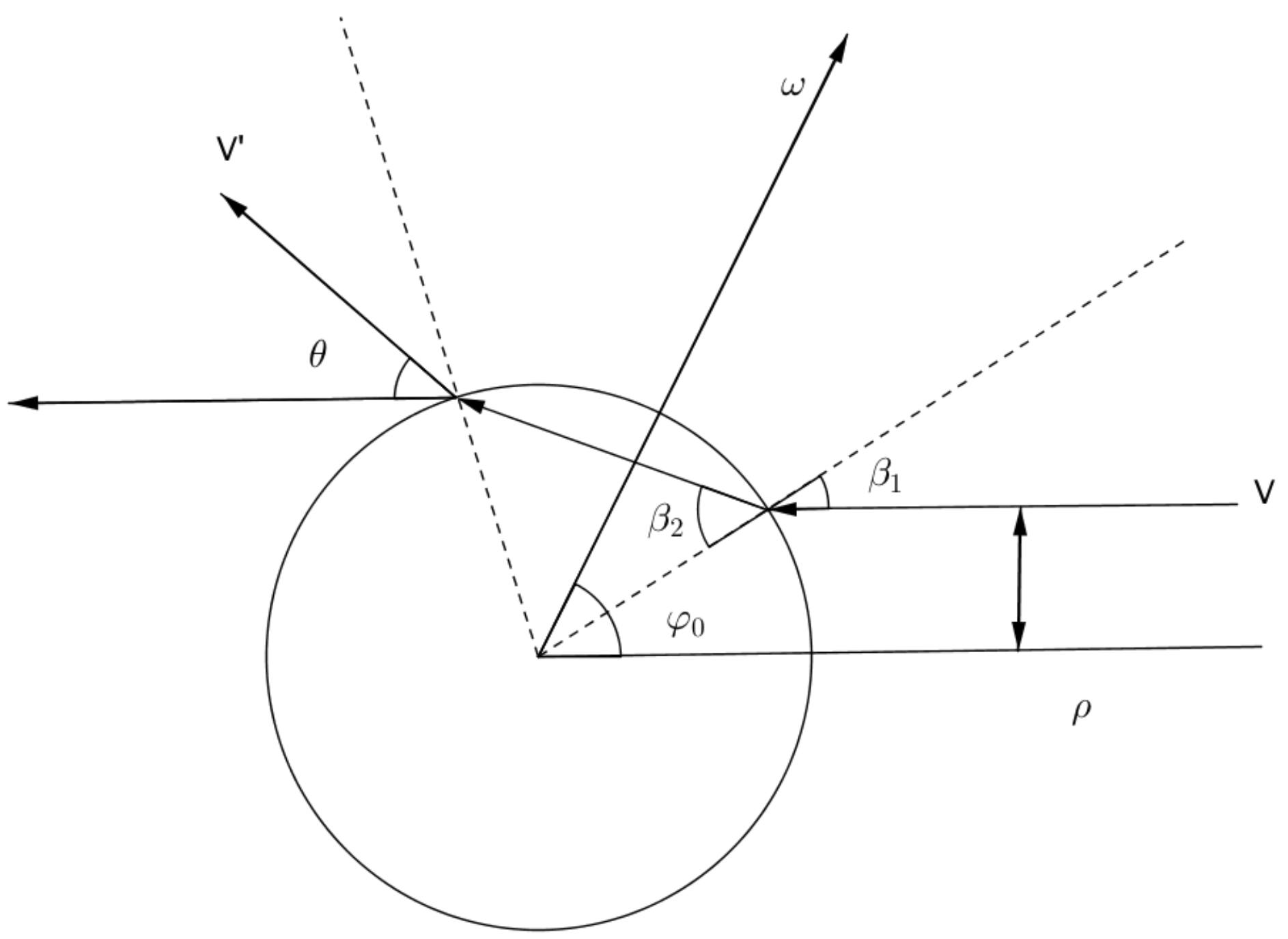}
\caption{Scattering by a spherical potential barrier. The particle moves in a straight line which is refracted on entering and leaving the barrier.}\label{fig:1}
\end{figure}

 Now let $T^t_{\mbf{c}_{N}}(x,v)$ be the Hamiltonian flow solution of Eq.n \eqref{scaled} with initial datum $(x,v)$ in a given sample $\mbf{c}_{N}=(c_1,\dots,c_N)$ of obstacles (skipping the $\ep$ dependence for notational simplicity). $T^t_{\mbf{c}_{N}}(x,v)$ is generated by the Hamiltonian
\begin{equation}\label{Hamilt}
H(x,v,\mbf{c}_{N})=\frac{1}{2}v^2+\ep^{\alpha}\sum_{j}\phi\left(\frac{|x-c_j|}{\ep}\right),
\end{equation}
where $\phi$ is given by \eqref{pot barr}. 
For this choice of the potential $\nabla\phi$ is not well defined. However the explicit solution of the equation of motion is obtained by solving the single scattering problem using the energy and angular momentum conservation (see Figure \ref{fig:1}). 

In Figure \ref{fig:1} we represent the scattering of a particle entering in the ball 
\begin{equation*}
B(0,1)=\{x\;\text{s.t.}\; |x|<1\}
\end{equation*} 
toward a potential barrier of intensity $\phi(x)=\ep^{\alpha}$.

Using the Snell law of refraction we have an explicit expression for the refractive index, i.e.
\begin{equation}
\label{refractive index }
n_{\ep}=\frac{\sin\beta_1}{\sin\beta_2}=\frac{|\bar{v}|}{|v|}=\sqrt{1-\frac{2\ep^{\alpha}}{v^2}},
\end{equation}
where $v$ is the initial velocity, $\bar v$ the velocity inside the barrier, $\beta_1$ the angle of incidence and $\beta_2$ the angle of refraction.
The scattering angle is $\theta=\pi-2\ffi_0=2(\beta_2-\beta_1)$ and the impact parameter is $\rho=\sin\beta_1$. 
 \begin{remark}Formula \eqref{refractive index } makes sense if $\frac{2\ep^{\alpha}}{v^2}<1$ and $\rho=\sin\beta_1<\sqrt{1-\frac{2\ep^{\alpha}}{v^2}}$.\\
When one of these two inequalities is violated, the outgoing velocity is the one given by the elastic reflection.
\end{remark}
A careful computation (see Appendix 1 in \cite{BNP} for further details) shows that the explicit expression for the scattering angle is given by 
\begin{equation}
\label{scattering angle ep}
\theta_{\ep}(\rho)=
\begin{cases}
2\left(\arcsin\left(\frac{\rho}{n_{\ep}}\right)-\arcsin(\rho)\right)  & \text{if } \rho \leq n_{\ep} \\
2\arccos(\rho)       & \text{if } \rho>n_{\ep}.
  \end{cases}
\end{equation}
Here we are not considering possible overlappings of obstacles. The scattering process can be solved in this case as well. However this event is negligible because of the moderate densities we are considering.

Now let $f_0=f_0(x,v)$ be the initial probability distribution. We are interested in characterizing the asymptotic behavior, under the scaling illustrated above, of the evolved distribution
\begin{equation}\label{valatt}
f_{\ep}(x,v,t)=\EE_{\ep}[f_0(T^{-t}_{\mbf{c}_{N}}(x,v))].
\end{equation}
We expect that the probability distribution \eqref{valatt}, in the limit $\ep\to 0$, solves a linear kinetic equation, more precisely the linear Landau equation.
However, due to the particular choice of the interaction potential, new features emerges at a mesoscopic level. The novelty, compared to \cite{DP} and \cite{K}, is that we have a logarithmic divergence of the diffusion coefficient appearing in the Landau equation, due to the lack of smoothness of the potential. This divergence suggests to look at a longer time scale in which a diffusion in space arises.
In fact, for a potential of the form \eqref{pot barr}, the classical formula giving the diffusion coefficient in the Landau equation \eqref{Landau}, i.e.
\begin{equation}\label{usualdiffcoeff}
B:=\lim_{\ep\to 0}{\frac{\mu\ep^{-2\alpha}}{2}|v|\int_{-1}^{1}{\theta_{\ep}^2(\rho)\,d\rho}},
\end{equation}
becomes
\begin{equation}\label{divdiffcoeff}
B=\lim_{\ep\to 0}\mu\left[\frac{2\alpha}{|v|^3}|\log(\ep)|\right]=+\infty,
\end{equation}
where $\theta_{\ep}$ is the scattering angle defined in \eqref{scattering angle ep}.
For the detailed computation of the diffusion coefficient we refer to \cite{BNP}, Appendix 2.  Roughly speaking, we can state that the asymptotic equation for the density of the Lorentz particle reads as
\begin{equation}\label{approxLandau}
(\partial_{t}+v\cdot\nabla_{x})f(x,v,t)\sim\,|\log\ep| \,\tilde{B}\, \Delta_{|v|} f(x,v,t), \quad \tilde{B}<+\infty.
\end{equation}
Hence, the asymptotic behavior of the mechanical system we are considering is the same as the Markov process ruled by the linear Landau equation with a diverging factor in front of the collision operator. This is equivalent to consider the limit in the Euler scaling of the linear Landau equation, which is trivial. Therefore we do not get any hydrodynamical equation and the system quickly thermalizes to the local equilibrium.
To detect something non trivial we have to look at a longer time scale $t\to |\log\ep|t$ in which the equilibrium starts to evolve. As expected, a diffusion in space arises.

The main results are summarized in the following theorem (\cite{BNP}, Theorem 2.1).
\begin{theorem}
\label{th:main th}
Suppose $f_0\in C_0(\R^2\times\R^2)$ a continuous, compactly supported initial probability density. Suppose also that $|D_{x}^kf_0|\leq C$, where $D_x$ is any partial derivative with respect to $x$ and $k=1,2$.
Assume $\mu_{\ep}=\ep^{-2\alpha-1}$, with $\alpha\in (0,1/8)$. Then the following statement holds 
\begin{equation}
\label{item1}
\lim_{\ep\to 0}f_{\ep}(x,v,t)=\langle f_0\rangle:=\frac 1{2\pi}\frac 1{|v|}\int_{S_{|v|}} 
{f_0(x,v)\,dv},
\end{equation}
$\forall t\in (0,T]$,  $T>0$.
The convergence is in $L^2(\R^2\times S_{|v|} )$.

Moreover, define $F_{\ep}(x,v,t):=f_{\ep}(x,v,t |\log\ep|)$. Then, for all $t\in [0,T)$, $T>0$, 
\begin{equation*}
\label{item3}
\lim_{\ep\to 0}F_{\ep}(x,v,t)=\rho(x,t), 
\end{equation*}
where $\rho$ solves the following heat equation
\begin{equation}
\label{item3b}
\left\{
 \begin{array}{l}\vspace{0.2cm}
\partial_t\varrho=D\Delta\varrho\\ 
\varrho(x,0)=\langle f_0\rangle, 
\end{array} \right.
\end{equation}
with $D$ given by the Green-Kubo formula
\begin{equation}\label{GK}
D=\frac{2}{\mu}|v|\int_{S_{|v|}}{v\cdot\big(-\Delta_{|v|}^{-1}\big)v\,dv}=\frac{2\pi}{\mu}|v|^2 \int_{0}^{\infty}\EE\big[v\cdot V(t,v)\big]\,dt, 
\end{equation}
where $V(t,v)$ is the stochastic process generated by $\Delta_{|v|}$ starting from $v$ 
and $\EE[\cdot]$ denotes the expectation  with respect to the invariant measure, namely the uniform measure on $S_{|v|}$.
The convergence is in $L^2(\R^2\times S_{|v|} )$.
\end{theorem}

\begin{remark}
To recover instead the kinetic picture, we can rescale suitably the density of the Poisson process. More precisely, if $\mu_{\ep}=\frac{\ep^{-2\alpha-1}}{|\log\ep|}$, the microscopic solution $f_{\ep}$ defined by \eqref{valatt} converges to the solution of the linear Landau equation \eqref{Landau} with a renormalized diffusion coefficient
$\displaystyle B:=\lim_{\ep\to 0}{\frac{\mu_\ep}{2}\ep\,|v|\int_{-1}^{1}{\theta_\ep^2(\rho)\,d\rho}}=\frac{2\alpha}{|v|^3}\mu.$
This is stated in \cite{BNP}, Theorem 2.2.  The explicit expression for the renormalized $B$ can be found in \cite{BNP}, Appendix 2.
\end{remark}

\vspace{4mm}
We have seen that, according to the particular choice for the potential we are considering, the natural divergence of the diffusion coefficient $B$ leads to a diffusion when we look at the system on a longer time scale. We can wonder if this result can be achieved in presence of a smooth, radial, short-range potential $\tilde{\phi}\in C^2([0,1])$. In the same spirit as in \cite{ESY}, \cite{KR}, we scale the variables, the density and the potential according to 
\begin{equation}\label{scaling 2}
\left\{
 \begin{array}{l}\vspace{0.2cm}
x\to\ep x\\ 
t\to\ep^{\lambda}\ep t,\\
\mu_{\ep}=\ep^{-(2\alpha+\lambda+1)}\mu\\
\tilde{\phi}\to\ep^{\alpha}\tilde{\phi}.
\end{array} \right.
\end{equation}
The naive idea is that the kinetic regime describes the system for kinetic times $O(1)$ (i.e. $\lambda=0$). One can go further to diffusive times provided that $\lambda>0$ is not too large. Indeed the distribution function $f_\ep$ ``almost'' solves
$$\left(\ep^{\lambda}\partial_t +v\cdot \nabla_x \right)f_\ep \approx \ep^{-2\alpha-\lambda} \text{L}_{\ep} f_\ep \approx
 \ep^{-\lambda}c\,\Delta_{|v|} f_\ep$$
which is the analogue of \eqref{approxLandau} above. 
In other words there is a scale of time for which
the system diffuses. However such times should not prevent the Markov property. This gives a constraint on $\lambda$. In fact, we can prove that 
there exists a threshold $\lambda_0=\lambda(\alpha)$, emerging from the explicit estimate of the set of pathological configurations producing memory effects,
s.t. for $\lambda<\lambda(\alpha)$, for $t>0$ and $\ep\to 0$,  
\begin{equation*}
f_{\ep}(x,v,\ep^{\lambda}t)\to \rho(x,t)\quad \text{in}\quad L^2(\R^2\times\R^2),
\end{equation*}
solution of the heat equation 
\begin{equation*}
\left\{
 \begin{array}{l}\vspace{0.2cm}
\partial_t\varrho=D\Delta\varrho\\ 
\varrho(x,0)=\langle f_0\rangle, 
\end{array} \right.
\end{equation*}
with $D$ given by the Green-Kubo formula
\begin{equation*}\label{GK2}
D=\frac{2}{\mu}|v|\int_{S_{|v|}}{v\cdot\big(-\Delta_{|v|}^{-1}\big)v\,dv}.
\end{equation*}
This is stated in \cite{BNP}, Theorem 6.1.

\section{Ideas of the Proof}
To give some ideas about how this machinery works we can divide the problem into two steps. 
The first step concerns the kinetic limit. We analyze how the limiting process, whose Fokker-Planck equation is the linear Landau equation, is obtained from the deterministic time evolution of the mechanical system in the intermediate regime described by Eq.n \eqref{scaling}. 
The other step shows how to pass from the kinetic description ruled by the linear Landau equation to the macroscopic picture where a diffusion in the position variable arises. We remind that the heat equation is, in the present case, the correct hydrodynamic equation.

\subsection{The Kinetic Description}\label{Kin Desc}
Following the explicit approach in \cite{G}, \cite{DR}, \cite{DP} we will show the asymptotic equivalence of $f_\ep$, defined by \eqref{valatt}, and $h_\ep$ solution of the following Boltzmann equation
\begin{equation}
\label{Boltzmann epsilon}
(\partial_t+v\cdot\nabla_x)h_{\ep}(x,v,t)=\text{L}_{\ep}h_{\ep}(x,v,t),
\end{equation}
where 
\begin{equation}
\text{L}_{\ep}h(v)=\mu\ep^{-2\alpha}|v|\int_{-1}^{1}\,d\rho\{h(v')-h(v)\}.
\end{equation}
Here $v'=v-2(\omega\cdot v)\omega$ where $\omega=\omega(\rho,|v|)$ is the unit vector obtained by solving the scattering problem associated to $\phi$ (see Figure \ref{fig:1}).
This allows to reduce the problem to the analysis of a Markov process which is an easier task. In fact, the series expansion defining $h_{\ep}$ (obtained perturbing around the loss term) reads as
\begin{equation}
\begin{split}
\label{formula6}
h_{\ep}(x,v,t)&=e^{-2\ep^{-2\alpha}|v|t}\sum_{Q\geq 0}\mu_{\ep}^{Q}\int_{0}^{t}dt_Q\dots\int_{0}^{t_{2}}dt_1\\&
\int_{-\ep}^{\ep}d\rho_1\dots\int_{-\ep}^{\ep}d\rho_Q\, f_0(\bar{\xi}_{\ep}(-t),\bar{\omega}_{\ep}(-t)).
\end{split}
\end{equation}
with
\begin{equation}\label{limiting trajectory}
\left\{\begin{array}{ll}
\bar{\xi}_{\ep}(-t)=x-vt_1-v_1(t_2-t_1)\dots-v_Q(t-t_Q)&\\
\bar{\omega}_{\ep}(-t)=v_Q.&
\end{array}\right.
\end{equation} 
We remark that $\bar{\omega}_{\ep}$ is an autonomous jump process and $\bar{\xi}_{\ep}$
is an additive functional of $\bar{\omega}_{\ep}$. Hence Eq.n \eqref{formula6} is an evolution equation for the probability density associated to a particle performing random jumps in the velocity variable at random Markov times. 

\vspace{2mm}
We consider the microscopic solution $f_\ep$ defined by  \eqref{valatt}. For $(x,v)\in\R^2\times\R^2$, $t>0$, we have
\begin{equation}
\label{formula1}
f_{\ep}(x,v,t)=e^{-\mu_{\ep}|B_t(x,v)|}\sum_{N\geq 0}\frac{\mu_{\ep}^{N}}{N!}\int_{B_t(x,v)^N}d\mbf{c}_{N}\, f_0(T^{-t}_{\mbf{c}_{N}}(x,v)),
\end{equation}
where $T^t_{\mbf{c}_{N}}(x,v)$ is the Hamiltonian flow generated by the Hamiltonian \eqref{Hamilt}. Here
$B_t(x,v):=B(x,|v|t)$, where $B(x,R)$ denotes the disk of center $x$ and radius $R$.
Thanks to suitable manipulations (see \cite{BNP} to go into details), 
Eq.n \eqref{formula1} becomes  
\begin{equation}
\begin{split}
\label{formula2}
f_{\ep}(x,v,t)=&\,\sum_{Q\geq 0}\frac{\mu_{\ep}^{Q}}{Q!}\int_{B_t(x,v)^Q}d\mbf{b}_{Q}\, e^{-\mu_{\ep}|\T(\mbf{b}_{Q})|}f_0(T^{-t}_{\mbf{b}_{Q}}(x,v))\\&\quad
\chi(\{\text{the}\;\mbf{b}_{Q}\;\text{are}\;\text{internal}\}),
\end{split}
\end{equation}
where here and in the sequel $\chi(\{\dots\})$ is the characteristic function of the event $\{\dots\}$ and the ``internal obstacles" are the obstacles of the configuration which, up to time $t$, influence the motion of the light particle. 
Moreover $\T(\mbf{b}_{Q})$ is the tube
\begin{equation}
\T(\mbf{b}_{Q})=\{y\in B_t(x,v)\;\text{s.t.}\;\exists s\in(-t,0)\;\text{s.t.}\;|y-x_{\ep}(s)|<\ep\}.
\end{equation}
Here $(x_{\ep}(s),v_{\ep}(s))=T^s_{\mbf{c}_{N}}(x,v)$.
We introduce
\begin{equation}
\begin{split}
\label{formula3}\displaystyle
\tilde{f}_{\ep}(x,v,t)=&\,e^{-2\ep^{-2\alpha}|v|t}\sum_{Q\geq 0}\frac{\mu_{\ep}^{Q}}{Q!}\int_{B_t(x,v)^Q}d\mbf{b}_{Q}\\&
\;\chi(\{\text{the}\;\mbf{b}_{Q}\;\text{are}\;\text{internal}\})\chi_1(\mbf{b}_{Q})f_0(T^{-t}_{\mbf{b}_{Q}}(x,v)),
\end{split}
\end{equation}
where $\chi_1$ is the characteristic function of the set of configurations $\mbf{b}_{Q}$ for which the particle is outside the range of all scatterers at time $0$ and at time $-t$, namely
\begin{equation}
\chi_1(\mbf{b}_{Q})=\chi\{\mbf{b}_{Q}\;\text{s.t.}\; b_i\notin B(x,\ep)\;\text{and}\; b_i\notin B(x_{\ep}(-t),\ep)\;\text{for}\;\text{all}\; i=1,\dots,Q\}.
\end{equation}
Since
\begin{equation}
|\T(\mbf{b}_{Q})|\leq 2\ep|v|t,
\end{equation}
we have
\begin{equation}
f_{\ep}\geq \tilde{f}_{\ep}.
\end{equation}

The key idea of this Markovian approximation is a suitable change of variables which has been introduced by Gallavotti in \cite{G}. 
We order the obstacles $b_1,\dots,b_Q$ according to the scattering sequence. Let $\rho_i$ and $t_i$ be the impact parameter and the entrance time of the light particle in the protection disk around $b_i$, i.e. $B(b_i,\ep)$. Hence we perform the following change of variables

\begin{equation}
\label{change var}
b_1,\dots,b_Q\rightarrow \rho_1,t_1,\dots,\rho_Q,t_Q,
\end{equation}
with
\begin{equation*}
0\leq t_1<t_{2}<\dots<t_Q\leq t.
\end{equation*}
Conversely, fixed the impact parameters $\{\rho_i\}$ and the hitting times $\{t_i\}$ we construct the centers of the obstacles $b_i=b(\rho_i,t_i)$. By performing the backward scattering we construct a trajectory $(\xi_{\ep}(s),\omega_{\ep}(s))$, $s\in[-t,0]$.
However $(\xi_{\ep}(s),\omega_{\ep}(s))=(x_{\ep}(s),v_{\ep}(s))$ (therefore the mapping \eqref{change var} is one-to-one) only outside the following pathological situations.\\
i) \textbf{Overlapping}.\\ If $b_i$ and $b_j$ are both internal and $B(b_i,\ep)\cap B(b_j,\ep)\neq\emptyset$ .\\
ii) \textbf{Recollisions}.\\ There exists $b_i$ such that for $\tilde{s}\in(t_{j},t_{j+1})$, $j>i$, $\xi_{\ep}(-\tilde{s})\in B(b_i,\ep)$. \\
iii) \textbf{Interferences}.\\ There exists $b_i$ such that $\xi_{\ep}(-\tilde{s})\in B(b_i,\ep)$ for $\tilde{s}\in(t_{j},t_{j+1})$, $j<i$.

\begin{figure}[htbp]
\includegraphics[scale=0.4]{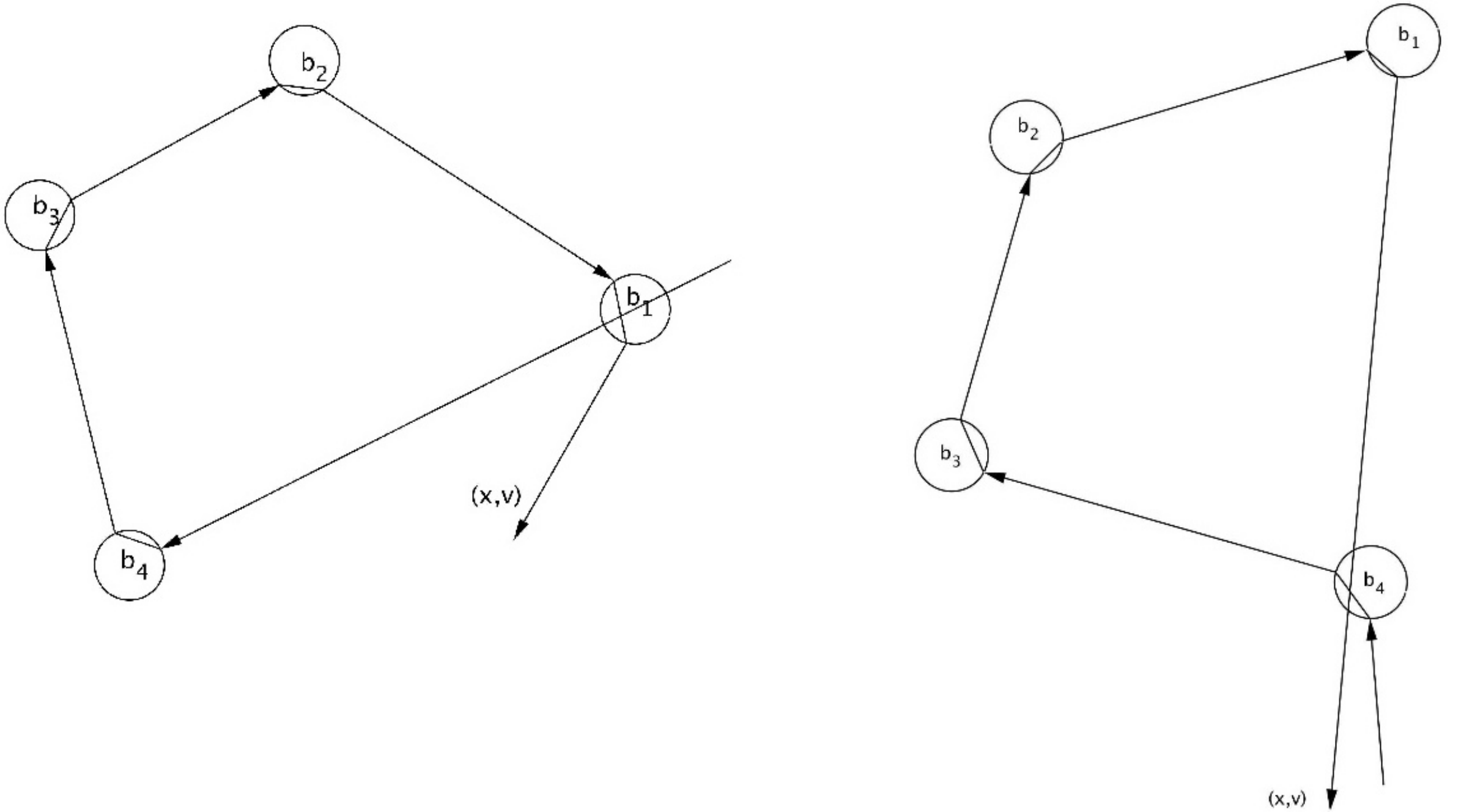}
\caption{Pathological events: on the left the backward trajectory delivers a recollision, namely the obstacle whose center is $b_1$ is recollided in the time interval $(-t_5,-t_{4}^- )$, 
on the right the backward trajectory delivers an interference, namely the obstacle whose center is $b_4$ belongs to the tube spanned by $\xi_{\ep}(-\tilde{s})$ for $\tilde{s}\in(0,t_{1})$. }
\label{fig:5}
\end{figure}

In order to skip such events we define
\begin{equation}
\begin{split}
\label{formula4}
\bar{f}_{\ep}(x,v,t)&=e^{-2\ep^{-2\alpha}|v|t}\sum_{Q\geq 0}\mu_{\ep}^{Q}\int_{0}^{t}dt_Q\dots\int_{0}^{t_{2}}dt_1\\&\quad
\int_{-\ep}^{\ep}d\rho_1\dots\int_{-\ep}^{\ep}d\rho_Q\,\chi_1(1-\chi_{ov})(1-\chi_{rec})(1-\chi_{int})f_0(\xi_{\ep}(-t),\omega_{\ep}(-t)),
\end{split}
\end{equation}
where $\chi_{ov}$, $\chi_{int}$ and $\chi_{rec}$ are the characteristic functions 
of the events i), ii), iii) respectively. Moreover we observe that $$\displaystyle \bar{f}_{\ep}\leq\tilde{ f}_{\ep}\leq f_{\ep}.$$ 

Next we remove $\chi_1(1-\chi_{ov})(1-\chi_{rec})(1-\chi_{int})$ by setting 
\begin{equation}
\begin{split}
\label{formula5}
\bar{h}_{\ep}(x,v,t)&=e^{-2\ep^{-2\alpha}|v|t}\sum_{Q\geq 0}\mu_{\ep}^{Q}\int_{0}^{t}dt_Q\dots\int_{0}^{t_{2}}dt_1 \\&\quad
\int_{-\ep}^{\ep}d\rho_1\dots\int_{-\ep}^{\ep}d\rho_Q\, f_0(\xi_{\ep}(-t),\omega_{\ep}(-t)).
\end{split}
\end{equation}
Since
\begin{equation}
\label{error}
1-\chi_1(1-\chi_{ov})(1-\chi_{rec})(1-\chi_{int}) \leq (1-\chi_1)+\chi_{ov}+\chi_{rec}+\chi_{int},\vspace{2mm}
\end{equation}
from \eqref{formula4} and \eqref{formula5} we have \vspace{2mm}
\begin{equation*}
\bar{f}_{\ep}(t)=\bar{h}_{\ep}(t)+\ffi_1(\ep,t).
\end{equation*}
with
\begin{equation}\label{def:ffi1}
\begin{split}
\ffi_1(\ep,t):=& \,e^{-2\mu_{\ep}\ep t} \sum_{Q\geq 0}(\mu_{\ep})^Q\int_{0}^{t}dt_Q\dots\int_{0}^{t_{2}}dt_1\int_{-\ep}^{\ep}d\rho_1\dots\int_{-\ep}^{\ep}d\rho_Q 
\\&
\, \{(1-\chi_1)+\chi_{ov}+\chi_{rec}+\chi_{int}\}f_0(\xi_{\ep}(-t),\omega_{\ep}(-t)).
\end{split}
\end{equation}
We observe that this is the crucial part. In the two-dimensional case the probability of those bad behaviors producing memory effects (correlation between the past and the present) is nontrivial.
To control the unphysical trajectories we need an explicit estimate of the set of bad configurations of the scatterers, in other words we have to estimate the error term $\ffi_1(\ep,t)$ 
showing that it is negligible in the limit. In \cite{BNP}, Section 5, we prove that 
$$\|\ffi_1(\ep,t)\|_{L^1}\to 0\quad \text{as}\quad \ep\to 0\quad \text{for\, all} \quad t\in[0,T].$$
Moreover the control of memory effects still holds for a longer time scale, namely
\begin{equation*}
\|\ffi_1(\ep,t)\|_{L^1}\underset{\ep\to 0}{\longrightarrow} 0\quad\forall t\in[0,|\log\ep|T],\quad T>0,
\end{equation*}
which implies that we can look at the system on a longer time scale.

\vspace{2mm}Since we are working to achieve the asymptotic equivalence of $f_\ep$ and $h_{\ep}$, we need to compare $\bar{h}_{\ep}$ with $h_\ep$. 
This is fulfilled once we consider the collision as instantaneous. More precisely, for the sequence $t_1,\dots,t_Q$ $\rho_1,\dots\rho_Q$ consider the sequence $v_1,\dots,v_Q$ of incoming velocities before the Q collisions. This allows to construct the limiting trajectory $\bar{\xi}_{\ep}(-t)$, given by \eqref{limiting trajectory}, which approximates the trajectory $\xi_{\ep}(-t)$ up to an error vanishing in the limit.
Indeed, due to the Lipschitz continuity of $f_0$, we can assert that
\begin{equation}
\bar{h}_{\ep}(x,v,t)=h_{\ep}(x,v,t)+\ffi_2(x,v,t),
\end{equation}
where 
\begin{equation}\label{phi2}
\sup_{x,v,t\in[0,T]}|\ffi_2(x,v,t)|\leq C\ep^{1-2\alpha}T.
\end{equation}

\subsection{The Diffusive limit}\label{Diff lim} 
For the sake of simplicity we set $\eta_\ep=|\log \ep|$.
We rewrite the linear Boltzmann equation \eqref{Boltzmann epsilon} in the following way
\begin{equation}\label{BE0}
\big(\partial_t +v\cdot\nabla_x\big)h_{\ep}(x,v,t)=\eta_\ep\,\tilde{ \text{L}}_{\ep} h_{\ep}(x,v,t),
\end{equation}
where $\tilde L_\ep=L/\eta_\ep$, namely
\begin{equation}\label{def:L_ve}
\tilde{ \text{L}}_{\ep}f (v)=\mu|v|\frac{\ep^{-2\alpha}}{|\log\ep|}\int_{-1}^1d\rho\big[f(v')-f(v)\big].
\end{equation}

Performing the limit $\eta_\ep\to \infty$ in \eqref{BE0} we get a trivial result (Eq.n. \eqref{item1}, Theorem \ref{th:main th}).
To obtain something non trivial we look at the solution for times $\eta_\ep t$, thus
performing the diffusive scaling for space and time. 

We denote by 
$\h:=h_\ep(x , v, \eta_\ep t)$ 
the solution of the following rescaled linear Boltzmann equation  
\begin{equation}\label{BE1}
\big(\partial_t  +\eta_\ep\,v\cdot\nabla_x\big)\h=\eta_\ep^2\, \tilde{ \text{L}}_{\ep}\h,
\end{equation}
and we introduce the rescaled Landau equation
\begin{equation}\label{rescLandau}
\big(\partial_t +{\eta_{\ep}}\, v\cdot\nabla_x\big)g_{\eta_{\ep}}(x,v,t)={\eta_{\ep}}^2  \LL g_{\eta_{\ep}}(x,v,t),
\end{equation}
where $\displaystyle \LL=\frac{\mu}{2}\frac 1 {|v|}\Delta_{|v|}$. 

Firstly we compare $g_{\eta_{\ep}}$ with $\h$ to show that they are asymptotically equivalent. We look at the evolution of $\h-g_{\eta_\ep}$, namely 
\begin{equation*}
\big(\partial_t+\eta_\ep\, v\cdot\nabla_x \big)\big(\h-g_{\eta_\ep})
=\eta_\ep ^2\Big(\tilde{\text{L}}_\ep\h-\LL g_{\eta_\ep} \Big),
\end{equation*}
and we obtain
\begin{equation}\label{energyest}
\displaystyle \frac 1 2 \,\partial_t \|\h-g_{\eta_\ep}\|\leq \eta_\ep^2\big\|\big(\tilde{\text{L}}_\ep-\LL \big)
g_{\eta_\ep}\big\|,
\end{equation}
where $\|\cdot\|$ denotes the $L^2-$norm.
A straightforward computation shows that the collisions are grazing: each collision changes only slightly the velocity of a particle (for the detailed analysis of the scattering angle \eqref{scattering angle ep} we refer to  \cite{BNP}, Appendix 1).
Therefore we perform a Taylor's expansion of $g_{\eta_{\ep}}$ with respect to the velocity variable and we get
\begin{equation*}
\begin{split}
\tilde{\text{L}}_\ep g_{\eta_\ep}&=\,\mu |v|\frac{\ep^{-2\alpha}}{|\log\ep|}\int_{-1}^1 d\rho\,
\big[g_{\eta_\ep}(x, v', t)-g_{\eta_\ep}(x,v,t)  \big] 
\\&=\, \mu |v|\frac{\ep^{-2\alpha}}{|\log\ep|}\big\{
\frac 1 2 \Delta_{|v|}g_{\eta_\ep}\int_{-1}^1 d\rho\, {|v'-v|^2}+\int_{-1}^1 d\rho\,\tilde{R}_{\eta_\ep}\big\},
\end{split}
\end{equation*}
with $\tilde{R}_{\eta_\ep}=O(|v'-v|^4)$.
Since 
\begin{equation*}
\lim_{\ep\to 0}\frac{\ep^{-2\alpha}}{|\log\ep|}\int_{-1}^1 d\rho\,|v'-v|^2=\lim_{\ep\to 0}\frac{\ep^{-2\alpha}}{|\log\ep|}\int_{-1}^1 d\rho\,\left(4\sin^2\frac{\theta_\ep(\rho)}{2}\right)=2\frac {\alpha}{|v|^4}
\end{equation*}
and
\begin{equation*}
\lim_{\ep\to 0}\frac{\ep^{-2\alpha}}{|\log\ep|}\int_{-1}^1d\rho\,
\tilde{R}_{\eta_\ep}=\lim_{\ep\to 0}\frac{\ep^{-2\alpha}}{|\log\ep|}\int_{-1}^1d\rho\, |v-v'|^4=\ep^\alpha|\log\ep|^\beta,\quad -1<\beta
<\frac  5 2 \alpha -1,
\end{equation*}
we have
$$\|\big(\tilde{\text{L}}_\ep-\LL \big)g_{\eta_\ep}\|\leq \ep^\alpha|\log\ep|^\beta\|\Delta^2_{|v|}g_{\eta_\ep}\| \leq C\ep^\alpha|\log\ep|^\beta, \quad\quad C>0,$$ 
for some $-1<\beta
<\frac  5 2 \alpha -1$ and $\ep$ sufficiently small. 
See \cite{BNP}, Appendix 2, for the details.
Consequently, using \eqref{energyest}, we have that $\h$, solution of \eqref{BE1}, is close to $g_{\eta_{\ep}}$, solution of \eqref{rescLandau}, in $L^2(\R^2\times S_{|v|})$.

\vspace{2mm}Therefore we need to prove that $g_{\eta_{\ep}}$, solution of \eqref{rescLandau}, converges to 
$\varrho$ as $\ep\to 0$. The convergence is in $L^2(\R^2 \times S_{|v|})$,
uniformly in $t\in(0,T]$. 
$\varrho:\R^2\times[0,T]\to\R_+$ is the solution of the diffusion equation \eqref{item3b}, i.e. 
\begin{equation*}
\left\{
 \begin{array}{l}\vspace{0.2cm}
\partial_t\varrho=D\Delta\varrho\\ 
\varrho(x,0)=\langle f_0\rangle, 
\end{array} \right.
\end{equation*}
where $\displaystyle \langle f_0\rangle=\frac 1 {2\pi}\frac 1 {|v|}\int_{S_{|v|}} f_0(x,v)\,dv\;$ and  
$\;\displaystyle D=\frac 2 {\mu} |v|\int_{S_{|v|}} v\cdot \big(-\Delta_{|v|}^{-1}\big)v\,dv.$

\vspace{3mm}
The proof relies on a classical tool which is the Hilbert expansion technique. 
The Hilbert expansion is a formal series, in powers of $\frac{1} {\eta_\ep}$, which allows to write $g_{\eta_{\ep}}$ in the following way
$$
g_{\eta_{\ep}}(x,v,t)=g^{(0)}(x,t)+\sum_{k=1}^{+\infty} \left(\frac{1}{\eta_\ep}\right)^k\, g^{(k)}(x,v,t),
$$
where the coefficients $g^{(k)}$ are independent of $\eta_\ep$. The well known idea is to determine them recursively, by imposing that $g_{\eta_{\ep}}$ is a solution of \eqref{rescLandau}.
For the complete statement and the detailed computations we refer to \cite{BNP}, Section 4.

We assume that the initial datum of the Cauchy problem associated to \eqref{rescLandau} depends only on the position variable, namely the initial datum has the form of a local equilibrium, i.e. $g_{\eta_{\ep}}(x,v,0)=\langle f_0\rangle$. Moreover we require $g^{(0)}$ to satisfy the same initial condition as the whole solution $g_{\eta_{\ep}}$, namely $g^{(0)}(x,0)=\langle f_0\rangle$.
We consider the truncated expansion for $g_{\eta_{\ep}}$ at order $\eta_\ep^{-2}$, namely
$$g_{\eta_{\ep}}(x,v,t)=g^{(0)}(x,t)+\frac 1 {\eta_{\ep}}g^{(1)}(x,v,t)+\frac 1{\eta_{\ep}^2}g^{(2)}(x,v,t)
+\frac 1 {\eta_{\ep}}R_{\eta_{\ep}},$$
where $g^{(i)}$, $i=0,1,2$ are the first three coefficients of a Hilbert expansion in $\eta_\ep$,
and $R_{\eta_{\ep}}$ is the reminder.
Comparing terms of the same order in $\eta_\ep$, in \eqref{rescLandau}, we obtain the following equations
\begin{equation*}\begin{split}
&(i)  \,v\cdot\nabla_x g^{(0)}=\frac{\mu} 2 \frac 1 {|v|}\,\Delta_{|v|} g^{(1)}\\
&(ii)  \,\partial_t\, g^{(0)}+v\cdot\nabla_x g^{(1)}=\frac{\mu} 2 \frac 1 {|v|}\,\Delta_{|v|}
 g^{(2)}\\
&(iii)  \,\big(\partial_t  +{\eta_{\ep}}\,v\cdot\nabla_x\big)R_{\eta_{\ep}}={\eta_{\ep}}^2 \frac{\mu} 2 \frac 1 {|v|}\,\Delta_{|v|} R_{\eta_{\ep}}
-A_{\eta_{\ep}}(t),
\end{split}\end{equation*}
with $A_{\eta_{\ep}}(t)=A_{\eta_{\ep}}(x,v, t)=\partial_t g^{(1)}+\frac 1 {\eta_{\ep}}\partial_t  g^{(2)}
+v\cdot\nabla_x g^{(2)}$.

From Eq.ns (i), (ii), thanks to suitable computations, we obtain that $g^{(0)}$ 
is the solution of the heat equation \eqref{item3b}. Hence, by showing 
that the coefficients $g^{(i)}\in L^2({\R^2\times S_{|v|}})$ and that $R_{\eta_{\ep}}$ is uniformly bounded in $L^2$, we have that $g_{\eta_{\ep}}$ converges to $g^{(0)}$ in $L^2$ as $\eta_\ep\to\infty$. 
We observe that we have assumed as initial condition for the linear Landau equation a local equilibrium state (independent of $v$). Strictly speaking this is not necessary since there is an initial regime (the initial layer) in which a general state, depending also on the velocity variable, thermalizes very fast in time and locally in space (see \cite{BNP}, Lemma 4.1).

\section{Perspectives: transport properties of the Lorentz Gas}\label{sec:Persp}
Roughly speaking, to consider the system out of equilibrium, we consider the Lorentz gas in a bounded region in the plane and couple the system with two mass reservoirs at the boundaries. 
We wonder about what to expect in a stationary non equilibrium state. 
The naive physical intuition tell us that there there exists a stationary state for which
\begin{equation}
\label{FL}
J \approx -D \nabla \rho 
\end{equation}
where $J$ is the mass current, $\rho$ is the mass density and $D>0$ is the diffusion coefficient. Formula \eqref{FL} is the well known Fick's law which we want to prove in the present context.

In \cite{BNPP} we deal with the validation of the Fick's law of diffusion for the following model. 
We consider the slice  $\Lambda= (0,L)\times \R$ in the plane. In the left half-plane $(-\infty, 0)\times \R$ there is a free gas of light particles at density  $\rho_1$ with correlation functions given by $$f_{j}^{1}(x_1,v_1\dots, x_j,v_j) =( \rho_{1})^{j}M(v_1)\dots M(v_j),\quad j\geq 1,$$ where $M(v_i)$ is the density of the uniform distribution on the unit circle $S_1$. 
In the right half-plane $ (L,+\infty)\times \R$ there is a free gas of light particles at density  $\rho_2$ with correlation functions given by $$f_{j}^{2}(x_1,v_1\dots, x_j,v_j) =( \rho_{2})^{j}M(v_1)\dots M(v_j)\quad j\geq 1.$$  The two half-planes play the role of mass reservoirs. 
Inside $\Lambda$ there is a Poisson distribution of hard core scatterers of diameter $\ep$ and intensity $\mu$. We denote by $c_1,\dots,c_N$ their centers.  

\begin{figure}[ht]
\centering
\includegraphics[scale= 0.5]{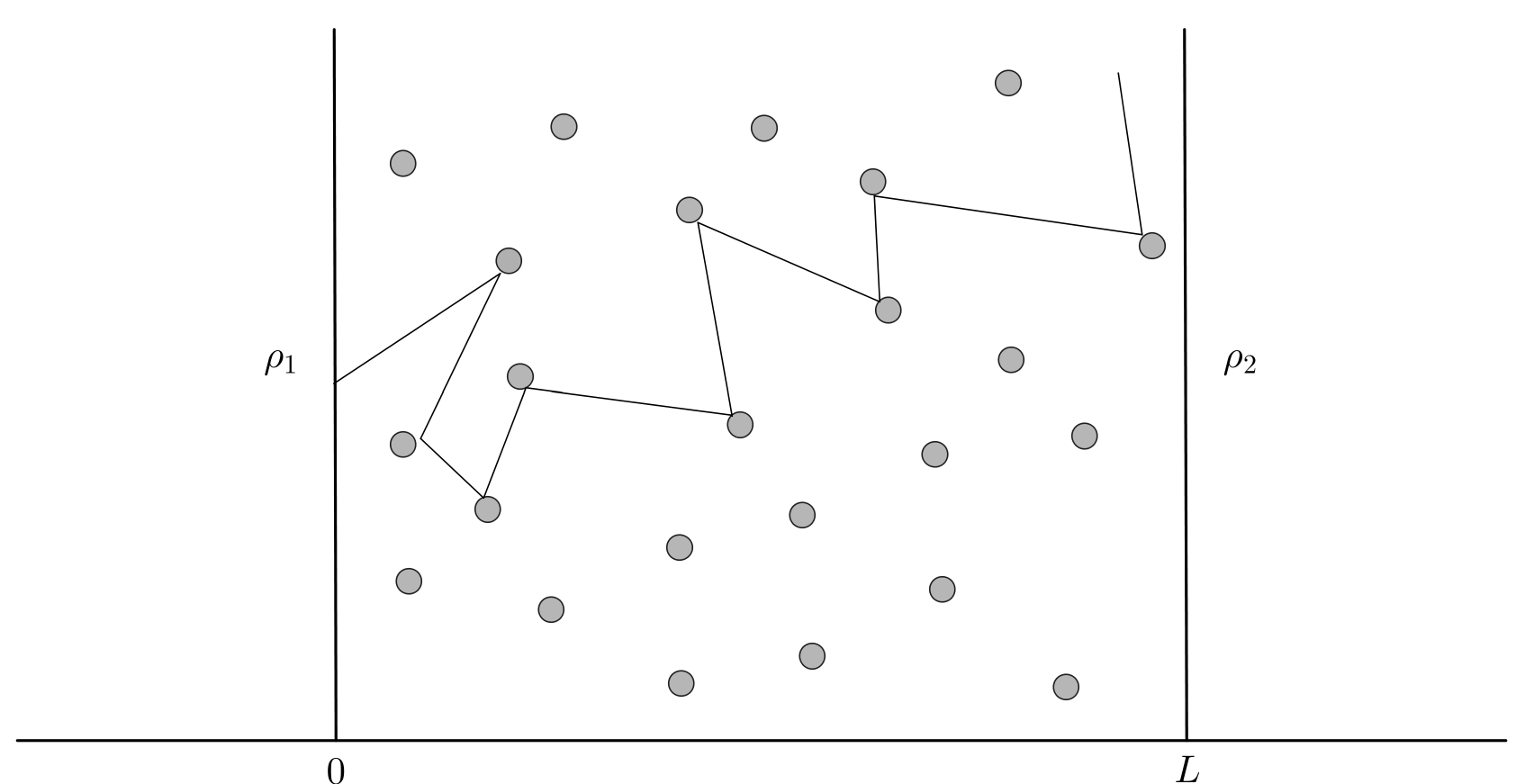}
\caption{$\Lambda$}
\label{fig:Transport}
\end{figure}

A particle in $\Lambda$ moves freely up to the first instant of contact with an obstacle. Then it is elastically reflected and so on. See Figure \ref{fig:Transport}.

Since the modulus of the velocity of the test particle is constant, we assume it to be equal to one, so that the phase space of our system is $\Lambda\times S_1$.

We rescale the intensity $\mu$ of the obstacles as $\mu_{\ep}=\mu\ep^{-1}\eta_{\ep}$
where, from now on, $\mu >0$ is fixed and $\eta_{\ep}$ is slowly diverging as $\ep\to 0$.  More precisely we assume that 
\begin{equation}\label{hp:etaep}
\ep^{\frac{1}{2}}\eta_{\ep}^{6}\to 0 \quad\text{as}\quad\ep\to 0.
\end{equation}
Hence we are in a low density regime and the scatterer configuration is dilute.

For a given configuration of obstacles $\mbf{c}_N$, we denote by $T^{-t}_{\mbf{c}_{N}}(x,v)$ the (backward) 
flow with initial datum $(x,v)\in\Lambda\times S_1$ and define
$t-\tau$, $\tau=\tau(x,v,t,\mbf{c}_N)$, as the first (backward) hitting time with the boundary. With $\tau=0$ we indicate the event such that the trajectory $T^{-s}_{\mbf{c}_{N}}(x,v)$, $s\in [0,t]$, never hits the boundary.
For any $t\geq 0$ the one-particle correlation function reads
\begin{equation}\label{def:fep}
f_{\ep}(x,v,t)=\EE_\varepsilon[f_B (T^{-(t-\tau)}_{\mbf{c}_{N}}(x,v))\chi(\tau>0)] + \EE_\ep[f_0 (T^{-t}_{\mbf{c}_{N}}(x,v))\chi(\tau=0)],
\end{equation} 
where $f_0\in L^\infty(\Lambda\times S_1)$ and the boundary value $f_B$ is defined by
\begin{equation}\label{def:fB}
f_B(x,v):=\left\{\begin{array}{ll}
\rho_1\quad\text{if}\quad x\in \{0\}\times\R,\quad v_1>0,&\vspace{3mm} \\
\rho_2\quad\text{if}\quad x\in \{L\}\times\R,\quad v_1<0,& \vspace{3mm}
\end{array}\right.
\end{equation}
with $\rho_1, \rho_2>0$. Here we absorbed $M(v)=\frac{1}{2\pi}$, the density of the uniform distribution on $S_1$, in the definition of the boundary values $\rho_1, \rho_2$.
To deal with the stationary regime we need to introduce the stationary solution of the problem, $f_\ep^S(x,v)$, which solves  
\begin{equation}\label{def:ST}
f_{\ep}^S(x,v)=\EE_\varepsilon[f_B (T^{-(t-\tau)}_{\mbf{c}_{N}}(x,v))\chi(\tau>0)] + \EE_\ep[f_ \ep^S(T^{-t}_{\mbf{c}_{N}}(x,v))\chi(\tau=0)]. 
\end{equation} 
We observe that $f_\ep^S$ depends on the space variable only through the horizontal component $x_1$ since it inherits this feature from the boundary conditions. 
Moreover we introduce the following observables  
\begin{equation}\label{def:j}
J_\ep^S(x)=\eta_{\ep}\int_{S_1}v\,f_\ep^S(x,v)\,dv,
\end{equation}
\begin{equation}\label{def:mass}
\varrho_\ep^S(x)=\int_{S_1}f_\ep^S(x,v)\,dv,
\end{equation}
i.e. the stationary mass flux and the stationary mass density respectively. We define $J_\ep^S$ as the total amount of mass flowing through a unit area in a unit time interval and we observe that, although in a stationary problem there is no typical time scale, the factor $\eta_{\ep}$ appearing in the definition of $J_\ep^S$, is reminiscent of the time scaling necessary to obtain a diffusive limit.

We are interested in the determining the existence and uniqueness of $f_\ep^S$ and its asymptotic behavior.  We can prove that  there exists a unique stationary solution for the microscopic dynamics which converges to the stationary solution of the heat equation, namely to the linear profile of the density. Moreover, in the same regime, the macroscopic current in the stationary state is given by the Fick's law, with the diffusion coefficient determined by the Green-Kubo formula.

The main results are summarized in the following theorems (\cite{BNPP}, Theorem 2.1 and Theorem 2.2)
\begin{theorem}\label{th:stat}
For $\ep$ sufficiently small there exists a unique $L^\infty$ stationary solution $f_\ep^S$ for the microscopic dynamics (i.e. satisfying \eqref{def:ST}). Moreover, as $\varepsilon \to 0$ 
\begin{equation}\label{convergenzaSTAZ}
f_\ep^S\rightarrow \varrho^S,
\end{equation}
where $\varrho^S$ is the stationary solution of the heat equation, namely the linear profile of the density 
\begin{equation}\label{statheat}
\varrho^S(x)= \frac{\rho_1(L-x_1)+\rho_2 x_1}{ L}.
\end{equation}
The convergence is in $L^{2}((0,L)\times S_1)$.
\end{theorem}

\begin{theorem}
\textbf{\textit{[Fick's law]}} \label{th:fick}
We have 
\begin{equation}\label{FickL}
J_\ep^S+D\nabla_x\varrho_\ep^S\to 0
\end{equation}
as $\ep\to 0$. The convergence is in $\mathcal{D}'(0,L)$ and $D>0$ is given by the Green-Kubo formula. 
Moreover
\begin{equation}\label{Jlim}
J^S=\lim_{\ep\to 0}J_\ep^S(x),
\end{equation}
where the convergence is in $L^2(0,L)$
and 
\begin{equation}\label{FICK}
J^S=-D\,\nabla\varrho^S=-D\,\frac{\rho_2-\rho_1}{L},
\end{equation}
where $\varrho^S$ is the linear profile \eqref{statheat}.
\end{theorem}
\vspace{4mm}
As expected by physical arguments, the stationary flux $J^S$ does not depend on the space variable. Furthermore the diffusion coefficient $D$ is determined by the behavior of the system at equilibrium and in particular it is equal to the diffusion coefficient for the time dependent problem.

Note that Theorem \ref{th:fick} is a straightforward consequence of Theorem \ref{th:stat}. We observe that in order to prove the convergence of the stationary solutions \eqref{convergenzaSTAZ} we can relax the hypothesis on $\eta_\ep$: it is enough to require $\eta_\ep$ such that $\ep^{\frac 1 2}\eta_{\ep}^5\to 0.$ To prove instead Fick's law we need something more: we require $\eta_\ep$ to satisfy \eqref{hp:etaep}.  

Our result holds in a low-density regime, hence we can use the linear Boltzmann equation as a bridge between our original mechanical system and the diffusion equation. 
We exploited this basic idea to prove Theorem \ref{th:main th} by using the linear Landau equation as intermediate level of description (see also \cite{ESY}, \cite{BGS-R} where the linear Quantum Boltzmann equation and the linear Boltzmann equation respectively have been used to obtain the heat equation from the particle system in different contexts).
This works once one has an explicit control of the error in the kinetic limit, which suggests the scale of times for which the diffusive limit can be achieved. This 
explains briefly why the above constraint on $\eta_{\ep}$ emerges.
Moreover, since we are using the kinetic picture as an intermediate level of analysis, the diffusion coefficient $D$ appearing in Eq.ns \eqref{Jlim}, \eqref{FICK}, is given by the Green-Kubo formula for the linear Boltzmann equation.

Here we have an additional difficulty since we have to deal with a stationary problem. The basic idea is that the explicit solution of the heat equation and the control of the time dependent problem allow us to characterize the stationary solution of the linear Boltzmann equation. This turns out to be the basic tool to obtain the stationary solution of the mechanical system which is the main object in our investigation.

We stress that, to handle the stationary problem, we characterize the stationary solutions in terms of Neumann series, rather than as the long time asymptotics of the time dependent solutions. This trick avoids the problem of controlling the convergence rates, as $t \to\infty$, with respect to the scale parameter $\ep $.

Also in this case the strategy consists in two steps. The first one 
allows the transition from Boltzmann to the diffusion equation. It is in the same spirit of the one performed in Section \ref{Diff lim}. 
This is the Markov part since we approximate the Brownian motion by the Markov jump process whose generator is the Linear Boltzmann collision operator. We refer to \cite{BNPP}, Section 4, for a more detailed discussion. Despite the technique is standard, we need an apparently new analysis in $L^\infty$, for the time dependent problem (needed for the control of the Neumann series) and a $L^2$ analysis for the stationary problem.

The transition from the mechanical system to the Boltzmann equation in a low density regime can be read instead as a Markovian approximation, in the same spirit of the one performed in Section \ref{Kin Desc}. This allows the transition from the nonmarkovian mechanical system to a Markov process. To reach a diffusive behavior on a longer time scale we need to estimate the set of pathological configurations which produce memory effects. Hence the constructive approach due to Gallavotti is complemented by an explicit analysis of the bad events preventing the Markovianity. 
The complete analysis is faced in \cite{BNPP}, Section 5.

\begin{acknowledgement}
I am indebted to G. Basile, F. Pezzotti and M. Pulvirenti for their collaboration and for the illuminating discussions.
\end{acknowledgement}

\end{document}